# Programmable spatial coherence tomography: diffraction-limited three-dimensional reflection imaging under modulated monochromatic illumination

Hervé Hugonnet[1,2], Jieun Choi[2,3], Gyoung Hwan Kim[2,4], Chulmin Oh[1,2], Jimin Cho[2,5], Chungha Lee[1,2], Su-Jin Shin[6], Sujin Park[7], Bon-Kyoung Koo[7], Wang-Yuhl Oh[2,4], Pilhan Kim[2,3,8], and YongKeun Park[1,2,9*]

[1] Department of Physics, Korea Advanced Institute of Science and Technology (KAIST), Daejeon 34141, Republic of Korea
[2] KAIST Institute for Health Science and Technology, KAIST, Daejeon 34141, Republic of Korea
[3] Graduate School of Medical Science and Engineering, KAIST, Daejeon 34141, Republic of Korea
[4] Department of Mechanical Engineering, KAIST, Daejeon 34141, Republic of Korea
[5] Graduate School of Stem Cell and Regenerative Biology, KAIST, Daejeon 34141, Republic of Korea
[6] Department of Pathology, Gangnam Severance Hospital, Yonsei University College of Medicine, Seoul, Republic of Korea
[7] Center for Genome Engineering, Institute for Basic Science, Daejeon 34126, Republic of Korea
[8] BioMedical Research Center (BMRC), KAIST, Daejeon 34141, Republic of Korea
[9] Tomocube Inc., Daejeon 34109, Republic of Korea
[*] Corresponding author: Y.K.P (yk.park@kaist.ac.kr)

**Abstract**

Depth sectioning in reflection microscopy has predominantly relied on temporal coherence gating. Here we show that volumetric reflection tomography at diffraction-limited resolution can be achieved under monochromatic illumination by engineering spatial, rather than temporal, coherence. In programmable spatial coherence tomography (PSCT), a sequence of pupil-coded illumination patterns with angular-spectrum diversity generates measurement redundancy enabling the system to calibrate itself, jointly retrieving aberrations, illumination profiles, and sample motion without guide stars or modal priors. We demonstrate label-free volumetric imaging of thick human tissues, organoids, frequency-resolved dynamic contrast, and high-resolution *in vivo* brain imaging through a cranial window. These results position PSCT as an alternative to temporal coherence based reflection imaging in complex biological systems.

**Introduction**

Reflection microscopy provides intrinsic structural contrast from refractive-index variations within biological tissue, without requiring fluorescent labels or genetic modification. Achieving volumetric imaging at cellular resolution in this modality, however, has long demanded broadband temporal coherence for axial sectioning. Optical coherence tomography (OCT) exploits this principle using a low numerical aperture and long wavelengths to achieve deep penetration[1]. High-numerical-aperture implementations, including full-field optical coherence microscopy (FF-OCM)[2–4], extend lateral resolution but require symmetric interferometric geometries and stringent path-length and dispersion matching[4–7]. These hardware demands limit the robustness and accessibility of high-resolution reflection tomography while causing aberration that degrade the experimentally achievable resolution.

Efforts to address these constraints have followed several directions. Reflection phase microscopy[8–12] has introduced monochromatic or narrowband illumination combined with spatial incoherence to reduce dispersion requirements. However, these approaches have typically retained strict symmetric interferometric layouts.

Other approaches tackle the aberration challenge. For example, adaptive optics, pupil filtering, or reflection matrix approaches have further sought to improve resolution[7,13–16]. Optical coherence refraction tomography, for example, reconstructs sample-induced refractive perturbations using volumetric displacement analysis and sample rotation to then provide a high-resolution compound reflection image[17]. Nevertheless, many such methods rely on spectral diversity, guide-star assumptions, or modal aberration models, leaving room for improvements in living specimens constrained by aberration and motion artifacts.

Here we introduce programmable spatial coherence tomography (PSCT) in which volumetric sectioning arises from engineered spatial diversity rather than temporal coherence. By dynamically modulating the illumination pupil under monochromatic excitation, we introduce redundancy in Fourier space sampling. This redundancy enables joint retrieval of illuminations and aberrations, enabling calibration of transfer-function, motion stabilization and refocusing. A simplified single-objective geometry replaces conventional symmetric interferometric architectures, eliminating path-length and dispersion constraints while maintaining high numerical aperture operation. Off-axis holography provides single-shot complex-field acquisition, and Fourier-domain inversion reconstructs the three-dimensional scattering potential.

We demonstrate label-free diffraction-limited volumetric imaging across a range of demanding biological conditions, including thick human tissues, long-term live hepatic organoids monitored over 48 hours, and *in vivo* mouse brain through a thinned-skull cranial window. Beyond structural imaging, the programmable illumination enables frequency-resolved dynamic contrast that reveals subcellular organelle dynamics and functional information obtained without exogenous labelling. These results establish programmable spatial coherence as a design principle for reflection microscopy, in which coherence is engineered rather than gated, and calibration emerges from the measurement itself.

## Results

### Programmable spatial coherence tomography

Whereas OCT achieves axial resolution by exploiting broadband temporal coherence, PSCT generates depth sectioning with pupil-plane coded illuminations.

The overall optical layout of PSCT is presented in figure 1a. First, a temporally coherent but spatially incoherent source must be generated. A monochromatic laser (532 nm, Cobolt Samba, HÜBNER Photonics) is rendered spatially incoherent using a rotating diffuser, and its angular spectrum is dynamically modulated at the pupil plane via a digital micromirror device display (DLP LightCrafter 6500, Texas instruments) (Fig. 1a-i). Then, the reference and sample fields are separated by a polarization plate beam splitter (#12-646, Edmund optics) (Fig. 1a-ii) acting as a controllable split ratio half mirror where the reference is directly transmitted to the camera while the sample beam is reflected toward the sample. Finally, the sample beam illuminates the sample through the tube and objective lenses (Olympus, LUMFLN60XW NA=1.1) and the reflected field interferes with the reference beam to form an off-axis hologram, which is recorded by a camera (Atlas10 ATX051S-MC, LUCID Vision Labs) (Fig. 1a-iii).

Volumetric reconstruction is performed in the Fourier domain (Fig. 1b-e). After off-axis hologram demodulation via spatial filtering, the three-dimensional scattering potential is recovered by deconvolution (see Reconstruction algorithm section). The data content distinction between PSCT and conventional reflection microscopy is illustrated in Fourier space (Fig. 1f). In OCT and FF-OCM, axial sectioning arises from temporal coherence. In contrast, PSCT expands lateral spatial-frequency support with a high numerical aperture objective. The large numerical aperture also provides axial section achieving high-resolution in all three spatial directions.

This shift from temporal to spatial coherence enables a simplified single-objective optical setup that replaces conventional Linnik and Mirau interferometers [4–6,18–20]. As a result, alignment tolerances of the reference polarized beam splitter are relaxed to approximately 100 μm, and sensitivity to dispersion mismatch is eliminated. Off-axis holography is achieved by a slight tilt of the polarized beam splitter, enabling single-shot complex-field acquisition without reducing numerical aperture or introducing additional optical elements [8,9,11,21]. The resulting monochromatic compact single-objective configuration could be packaged in a unit compatible with standard inverted microscope platforms.

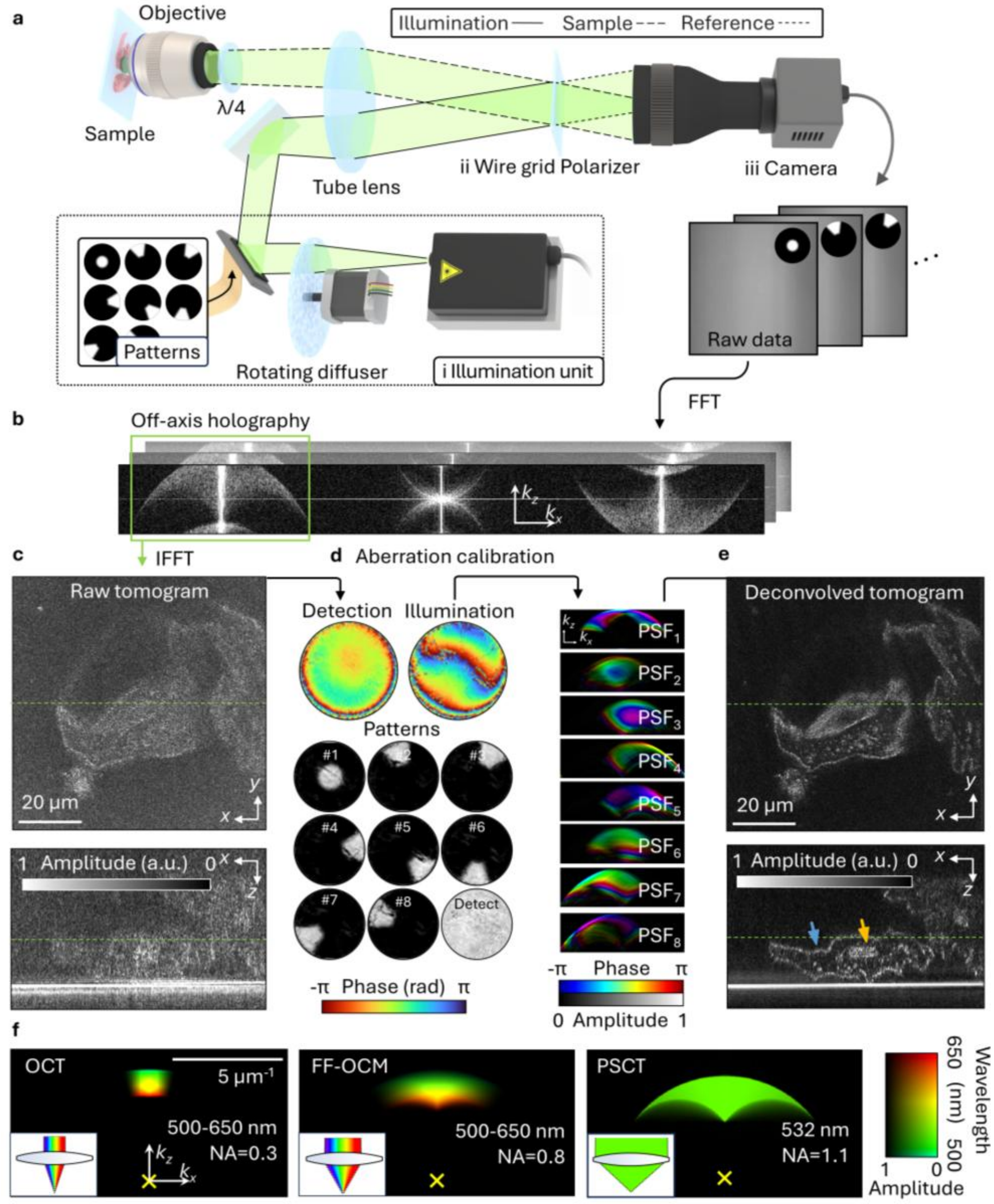

**Figure 1 | Programmable spatial coherence tomography.** **a** Optical setup with (i) the illumination unit (ii) the reference wire grid polarizer and (iii) the camera. **b** Fourier filtering for off-axis holography, **c** Raw reflection data, **d** self-calibration of aberrations and illumination are used to **e** reconstruct the hologram with deconvolution. Yellow arrow is the cell nucleus while the blue arrow indicates the cell membrane. **f** Spatial spectrum coverage of different reflection microscopes, PSCT achieves increased spectrum coverage in all x-y-z directions.

**Reconstruction algorithm**

The volumetric reconstruction recovers the three-dimensional scattering potential from a set of pupil-coded measurements. The current formulation assumes weak scattering under the first Born approximation, which is appropriate for the mesoscopic-to-cellular imaging regime (penetration depth <100 μm). Under spatially incoherent illumination for the pattern number $i$, each measurement $S_i(\mathbf{r})$ is the convolution of the scattering potential $F(\mathbf{r}) = k_0^2[n^2(\mathbf{r}) - n_m^2]$ with a point spread function $\mathrm{PSF}_i(\mathbf{r})$[22,23].

$$S_i(\mathbf{r}) = F(\mathbf{r}) * \mathrm{PSF}_i(\mathbf{r}),$$

with the sample refractive index $n(\mathbf{r})$, immersion index $n_m$, and wavevector $k_0 = n_m/\lambda$. For spatially incoherent reflection imaging, the effective point spread function (PSF) is given by the elementwise product of illumination and detection responses [24–26]:

$$\mathrm{PSF}_i(\mathbf{r}) = -16\pi^2 G_{\mathrm{ill},i}(\mathbf{r}) G_{\mathrm{det}}(\mathbf{r}),$$

where $G_{\mathrm{ill},i}$, $G_{\mathrm{det}}$ denote the illumination and detection Green's functions under programmed pupil modulation. The Green's functions are easiest to express in the Fourier space, varying with both optical aberrations $\phi(\mathbf{k})$ and pupil-plane illumination patterns $\rho(\mathbf{k})$.

$$G_{\mathrm{det}}(\mathbf{k}) = e^{j\phi_{\mathrm{det}}(\mathbf{k})}, G_{\mathrm{ill},i}(\mathbf{k}) = \sqrt{\rho_i(\mathbf{k})}e^{j\phi_{\mathrm{ill}}(\mathbf{k})}.$$

Here, $\rho_i(\mathbf{k})$ denotes the programmed pupil intensity pattern of illumination $i$ (Fig. 1ai), and $\phi_{\mathrm{det}}$ and $\phi_{\mathrm{ill}}$ represent detection and illumination aberration phases. Given measurements across multiple illumination conditions, the scattering potential is estimated in the Fourier space via a Tikhonov-regularized inversion[23]:

$$F_{recon}(\mathbf{k}) = \frac{\sum_i PSF_i^*(\mathbf{k})\, S_i(\mathbf{k})}{\sum_i |PSF_i(\mathbf{k})|^2 + \varepsilon},$$

Here $\varepsilon$ is fixed to approximately half the mean transfer-function intensity to prevent numerical instability in regions of low magnitude.

**Self-calibration and Aberration correction**

A key consequence of the programmable illumination design is that the redundancy of information across pupil patterns enables computational self-calibration of the point spread function parameters $\rho_i(\mathbf{k})$, $\phi_{\mathrm{det}}$ and $\phi_{\mathrm{ill}}$. By enforcing consistency between the convolution model and the measured data across all programmed illumination conditions. The self-calibration problem is formulated as:

$$\min_{\phi_{\mathrm{det}}, \phi_{\mathrm{ill}}, \{\rho_i\}} \left\{ \sum_i \int_{\mathbf{r}} \|S_i(\mathbf{r}) - F_{recon}(\mathbf{r}) * \mathrm{PSF}_i(\mathbf{r})\|_2^2 d\mathbf{r} + \int_{\mathbf{k}} \beta \left\| 1 - \sum_i \rho_i(\mathbf{k}) \right\|^2 d\mathbf{k} \right\}.$$

The first term enforces data fidelity under the forward model, while the second term weakly constrains the total pupil illumination to remain spatially homogeneous. Although not strictly required, this constraint improves convergence for challenging datasets, particularly *in vivo*. In this work, we set $\beta = 100$ and $\varepsilon = 10^{-12}$ in $F_{recon}$. Optimization was implemented in MATLAB using a custom automatic differentiation tool[27] and the Adamax optimizer[28]. Convergence was typically achieved within approximately 40 iterations.

In live samples, physiological motion complicates the deconvolution and self-calibration process and needs to be corrected to obtain noise free tomograms. At each axial position $z$, depth-dependent physiological motion parameters $\Delta x(z)$, $\Delta y(z)$, and $\Delta z(z)$ were estimated by maximizing the spectral overlap between the measured data $S_i(\mathbf{k}, z)$ and the non-aberrated point spread function $PSF_i(\mathbf{k})$. Lateral displacements $\Delta x(z)$ and $\Delta y(z)$ were modeled as phase ramps in Fourier space, while axial displacements $\Delta z(z)$ were modeled with a global phase shift $\theta(z) = 2k_0 \Delta z(z)$.

$$\max_{\Delta x(z), \Delta y(z), \theta(z)} \left\{ \int_{\mathbf{k}} \int_z \sum_i \left\| S_i(\mathbf{k}, z)\, PSF_i^*(\mathbf{k})\, e^{-j(k_x \Delta x + k_y \Delta y + \theta)} \right\|^2 dz d\mathbf{k} \right\}.$$

Optimization was performed using Nesterov accelerated gradient descent[29]. In contrast to interferometric synthetic aperture microscopy[7,15,16] and most computational adaptive optics approaches[13,14], the PSCT jointly estimates illumination and detection pupil functions together with illumination calibration and motion stabilization. Furthermore, retrieved aberrations are freeform without modal (e.g., Zernike) restriction. Estimating aberrations for both illumination and detection can correct for more generalized misalignments. For example, in figure 1e, the reference wire grid polarizer (Fig. 1a-ii) was misaligned by approximately 1mm resulting in a phase ramp difference between the illumination and detection aberrations. This misalignment was later resolved and aberration maps for other measurements are similar for detection and illumination. Deconvolution and aberration correction significantly improve resolution and reduce noise (Fig. 1e). After correction, the cell membrane (Fig. 1e blue arrow) and nucleus (Fig. 1e yellow arrow) are now clearly visible.

**Resolution validation**

Next, we evaluate PSCT's experimental resolution by imaging a multilayer scattering phantom composed of stacked adhesive tape layers (Fig. 2a). The refractive index mismatch between the tape ($n \approx 1.48$) and the water immersion induces increasing spherical aberration with imaging depth, leading to progressive loss of axial resolution and lateral contrast in the raw tomograms.

The recovered pupil aberration maps (Fig. 2b) correct this effect, recovering gradually increasing spherical aberrations with depth. The recovered aberration profiles correspond well to expected theoretical aberrations calculated for an RI of 1.48 ( cellulose RI=1.47 and binder RI=1.52 [30]) at depths of 33 and 67 μm (Fig. 2b), confirming accurate aberration recovery.

After aberration correction (Fig. 2c) resolution is maintained at every depth and scatterers within the tape are now well resolved. Aberration correction also increases data overlap across illumination patterns (Fig. 2d), indicating improved consistency of the measurements to the common sample.

Because the embedded scatterers are nanoscopic, their images approximate the effective point spread function, enabling quantitative resolution benchmarking. Comparison of raw and corrected reconstructions (Fig. 2e) shows the PSF narrowing to diffraction limited size. The full width at half maximum (FWHM) improves from 443 × 432

× 1956 nm$^3$ to 249 × 238 × 876 nm$^3$ after correction, approaching the theoretical diffraction limit of 230 × 230 × 810 nm$^3$ as calculated from the Fourier transform of the optical transfer function (Fig. 1f).

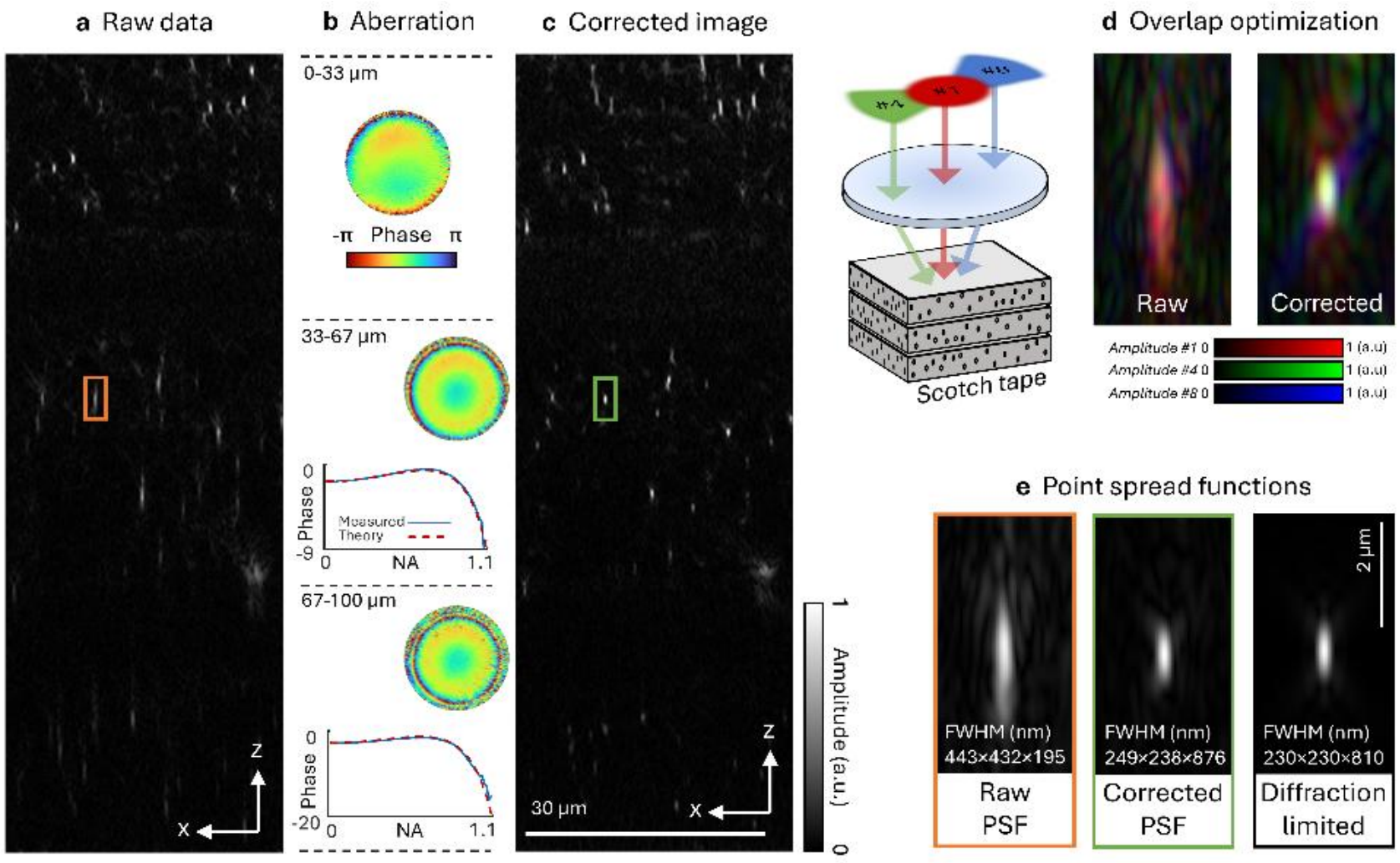

**Figure 2 | Computational aberration retrieval and diffraction-limited restoration. a** Raw tomograms acquired through stacked tape layers exhibit progressive axial and lateral blurring with imaging depth. **b** Retrieved aberrations increase in curvature with depth. Radial phase profiles (solid) agree with theoretical predictions (dashed). **c** Aberration-corrected reconstruction: self-calibration restores resolution and better resolves nanoscopic scatterers. **d** Overlap comparison of the signal from the different patterns before and after correction. **e** Point spread functions analysis shows near diffraction limited resolution after correction.

### Imaging of complex biological specimens

Having established near-diffraction-limited resolution, we next assessed whether PSCT can effectively image diverse biological specimens without fluorescent labels (Fig. 3). We apply PSCT to a large-area histological tissue section (Fig. 3a), a fixed hepatic organoid (Fig. 3b), and live mouse oocytes (Fig. 3c).

Large-area volumetric imaging of colon cancer tissue (Fig. 3a) shows the tissue organization into muscularis propria, submucosa and mucosa regions. In each region, a subregion of interest (ROI) (Fig. 3a i–iii) highlight distinct features. In the muscularis propria the ROI (Fig. 3ai) displays aligned muscle fibers. In the submucosa the ROI (Fig. 3aii) renders a vascular lumen. Finally, the ROI (Fig. 3aiii) in the mucosa displays comparatively isotropic morphology. Corresponding axial cross-sections confirm consistent depth sectioning throughout the imaging volume.

In fixed mouse hepatic organoids (Fig. 3b), PSCT resolves lumen structures and individual cells in three dimensions. In each cell the nucleus is distinctly visible in dark due to low light scattering. The orthogonal sections reveal that the sample is composed of three organoids stacked over each-over.

Finally, we evaluated PSCT applicability to live biological samples by imaging mouse oocytes (Fig. 3c). The three-dimensional reconstructions resolve subcellular organization, including the zona pellucida, nuclear apparatus and polar body (Fig. 3c, red, pink and blue arrows). The nuclear region exhibits reduced scattering relative to the surrounding cytoplasm, consistent with the organoid imaging result (Fig 3b).

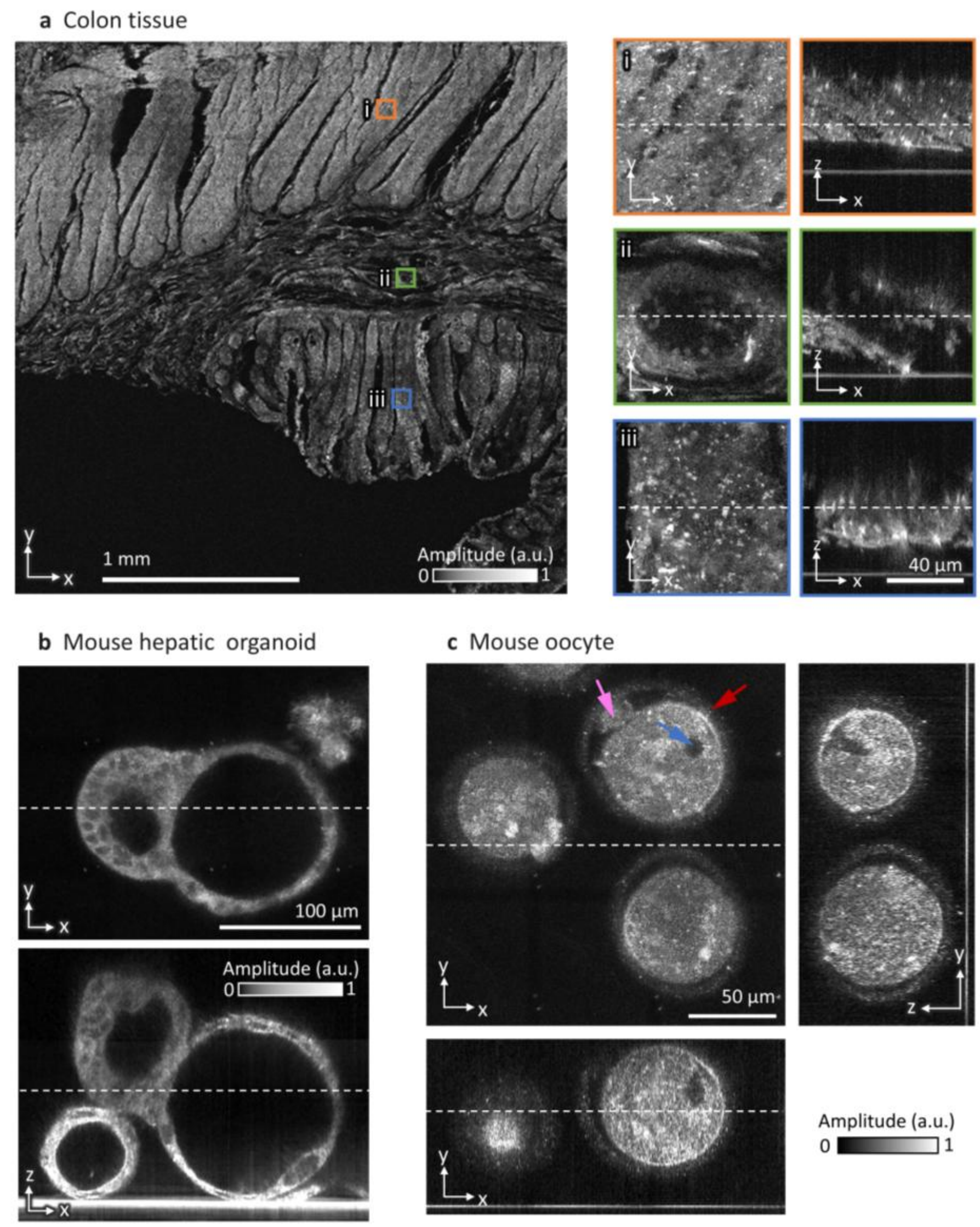

**Figure 3 | Label-free volumetric imaging of thick biological specimens. a** Large-area 3D tomogram of a human colon cancer tissue. Insets (i–iii) show representative regions with corresponding axial cross-sections (dashed lines). **b** Mouse hepatic organoids. **c** Live mouse oocytes; red, pink and blue arrows denote the zona pellucida, nuclear apparatus and polar body.

### Long-term live organoid imaging

A key advantage of the simplified single-objective geometry and computational self-calibration is compatibility with long-term live imaging, where environmental perturbations accumulate over time. To test this, we imaged live hepatic organoids over 48 hours at 12-hour intervals (Fig. 4).

In a healthy organoid (Fig. 4a), volumetric imaging resolves lumen growth, cell boundaries, and nuclei throughout the imaging period. Cells gradually thicken in the radial direction and pack more densely in the tangential direction. Cell polarity becomes apparent as the organoid matures: nuclei (lower-scattering structures) organize near the outer surface while the cytoplasm faces the lumen.

In an organoid with abnormal morphology (Fig. 4b), lumen formation initiated at the first time point was followed by rapid collapse and formation of a compact, dense structure. Individual cell volumes were larger than in the healthy organoid. While the sample size (1 sample per condition) precludes statistical conclusions, these observations demonstrate that PSCT provides sufficient volumetric contrast to resolve morphological differences between developmental phenotypes in live three-dimensional culture, motivating follow-up studies.

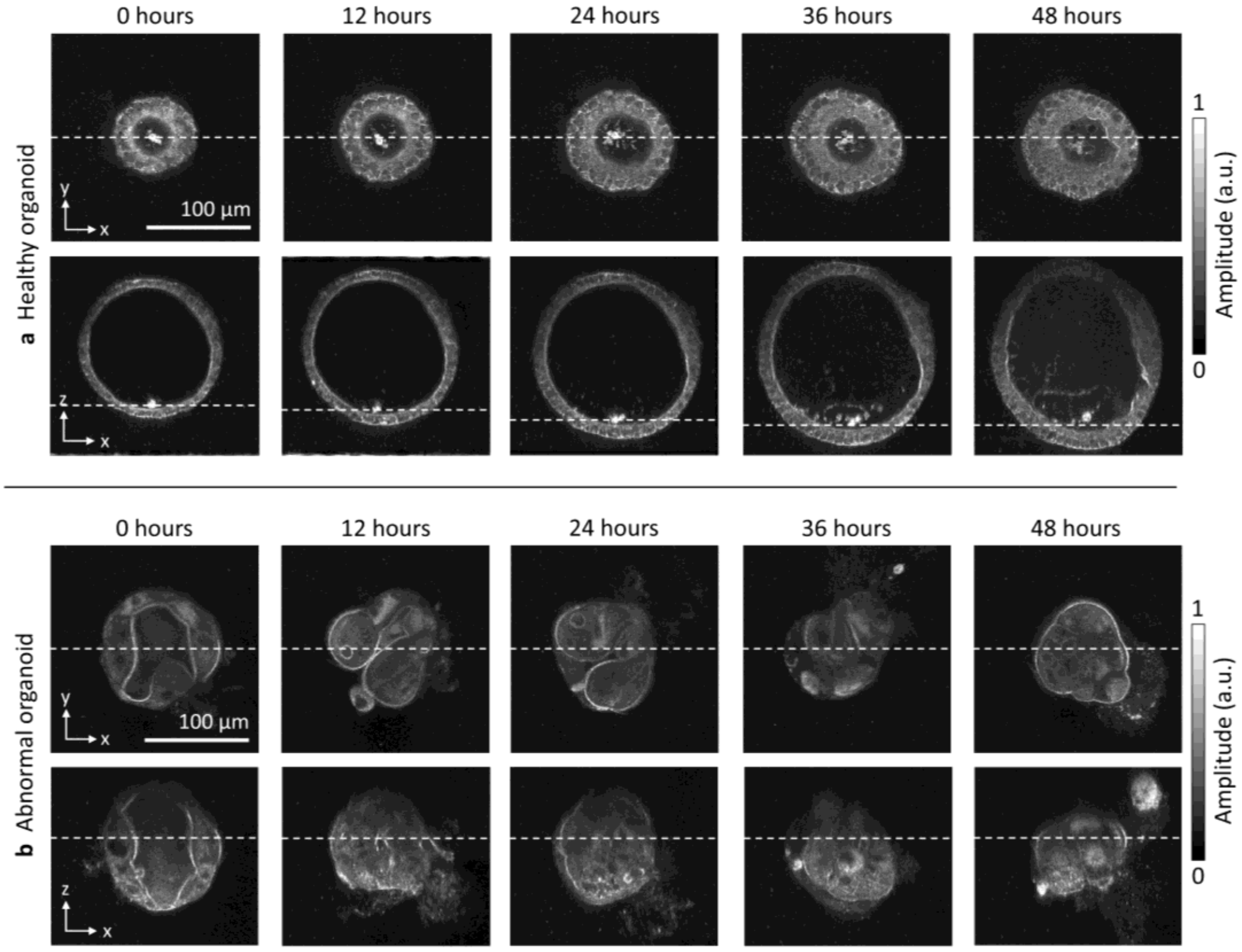

**Figure 4 | Long-term volumetric imaging of hepatic organoids. a** A healthy organoids grows over a 48-hours period. **b** A sick organoid fails to form a proper lumen structure.

**Volumetric dynamic contrast**

Dynamic imaging improves the resolution and specificity of reflection microscopy and yields promising results for various OCT configurations[4,31,32]. In conventional dynamic OCT, temporal fluctuations are analyzed at a single focal plane, which slows data acquisition. In PSCT, the programmed illumination enables simultaneous acquisition of data at several focal planes using computational refocusing. Indeed, focusing is expressed as a quadratic phase aberration at the pupil plane, and can be corrected within the PSCT reconstruction to generate data estimates at multiple depth from the data acquired at a single focal acquisition.

Exploiting this refocusing capability we perform dynamic imaging by reducing the scanning step from 500 nm to 16 nm and refocused over 4800 nm to form 300 times resolved tomogram recording 15 seconds of dynamics.

$$F(k_{xy}, z, t) \approx \sum_i S_i(k_{xy}, z, t_0) \mathrm{norm}[\mathrm{PSF}_i^*(k_{xy}, z + (t - t_0)v)]$$

Where $v$ is the axial scanning speed and $\mathrm{norm}[\cdot]$ is a function normalizing the amplitude to one except for amplitudes smaller than the one tenth of the maximum amplitude which are set to zero. This strategy requires approximately 10 times less time than imaging 300 dynamics independently at each depth. After stabilization and apodization with a Blackman window, the time resolved tomograms are analyzed based on the standard deviation within defined frequency bins (0-0.25Hz; 0.25-1.5Hz; 1.5-10Hz) and images are assigned to the blue green and red channels (Fig. 5).

The dynamic tomogram resolves subcellular compartments based on their characteristic motility (Fig. 5b). The tricellular junctions on the outer wall of the organoid show both low, medium and high frequency signals and are visible in white at cell’s corners (Fig. 5, ROI i). Other parts of the cell membranes exhibit minimal dynamics and are rendered deep blue. Medium and high frequency signals highlight subcellular components. The cytoplasm is dominated by medium frequency activity and is rendered green while the nucleus has faster activity rendering it brown. This contrast visualizes cell polarization: nuclei face the outer surface of the organoid while cytoplasm and organelles are positioned toward the lumen. Nucleus morphology is further highlighted by dark-looking nuclear envelopes and nucleoli. A cell undergoing mitosis which has two nuclei (Fig 5, ROI ii) also displays overall increased dynamics inside the cytoplasm making it pop up as bright green in the vertical cross-section. Finally, the lumen is rendered red due to fast dynamics, attributable to small cell debris and reduced viscosity compared with the cytoplasm[33]. Distinct homogeneous fluid compartments within the lumen, rendered black, appear to form by fusion of small vesicles emitted by the cells (Fig. 5, ROI iii).

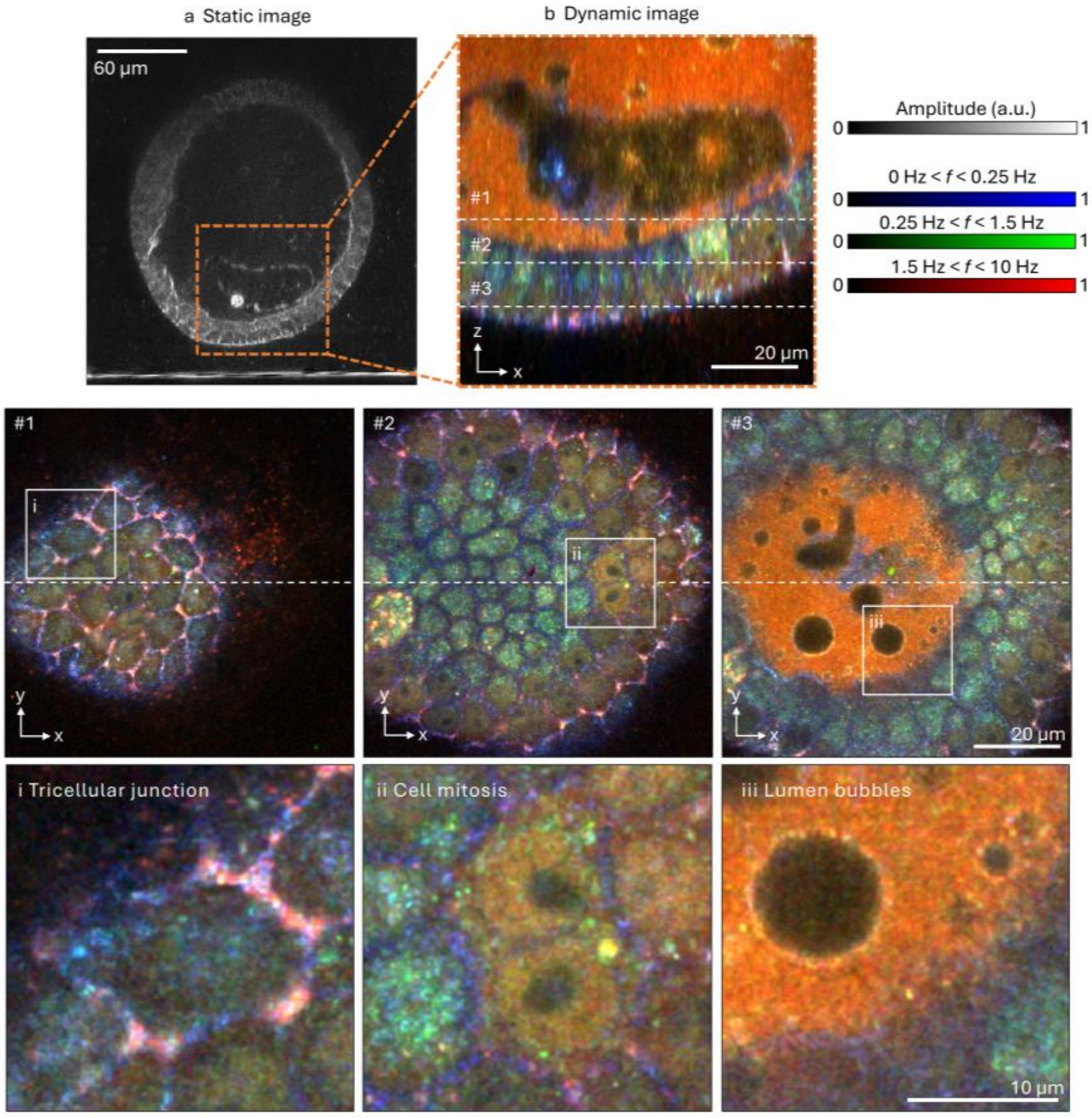

**Figure 5 | Organoid volumetric dynamic contrast. a** Static tomogram and **b** dynamic tomogram. Three cross-sections of the tomogram are shown with region of interest (i-iii). (i) Tricellular junction, (ii) cell mitosis and (iii) lumen bubbles.

**Mouse brain imaging**

The most stringent test for any reflection imaging system is *in vivo* operation, where refractive-index heterogeneity, sample tilt, and physiological motion occur simultaneously. We evaluated PSCT by imaging the mouse brain through a thinned-skull cranial window (Fig. 6a), a preparation that challenges all three aspects of the self-calibration framework: aberration correction, illumination retrieval, and motion stabilization.

Raw data (Fig. 6b) suffers from spherical and coma aberrations together with periodic axial displacement corresponding to respiratory motion (Fig. 6c). Motion was more pronounced in cortical tissue than in the skull, as the skull is fixed while the brain moves freely within. Aberration and motion correction restored axial resolution and structural contrast (Fig. 6d).

We stitch 5 × 5 tomograms to extend the field of view to 350 × 350 × 100 µm$^3$ (Fig. 6a). Cross-sections through the skull (Fig. 6e) resolve osteocytes within the bone matrix. Deeper, the fibrous structure of the dura and brain vasculature are clearly visible (Fig. 6f); blood vessels appear dark because the rapid blood motion averages the interferometric signal out. Vessel positions correspond well with the red CD31 fluorescence channel in the confocal fluorescence image (Fig. 6g). Finally, inside the cortex, myelinated axons are visible as bright linear structures in the tomogram (Fig. 6h), consistent with the high reflectance of myelinated structures as well as the general position of Thy1-GFP green fluorescence which is expressed in some of the mouse's neurons (Fig. 6i).

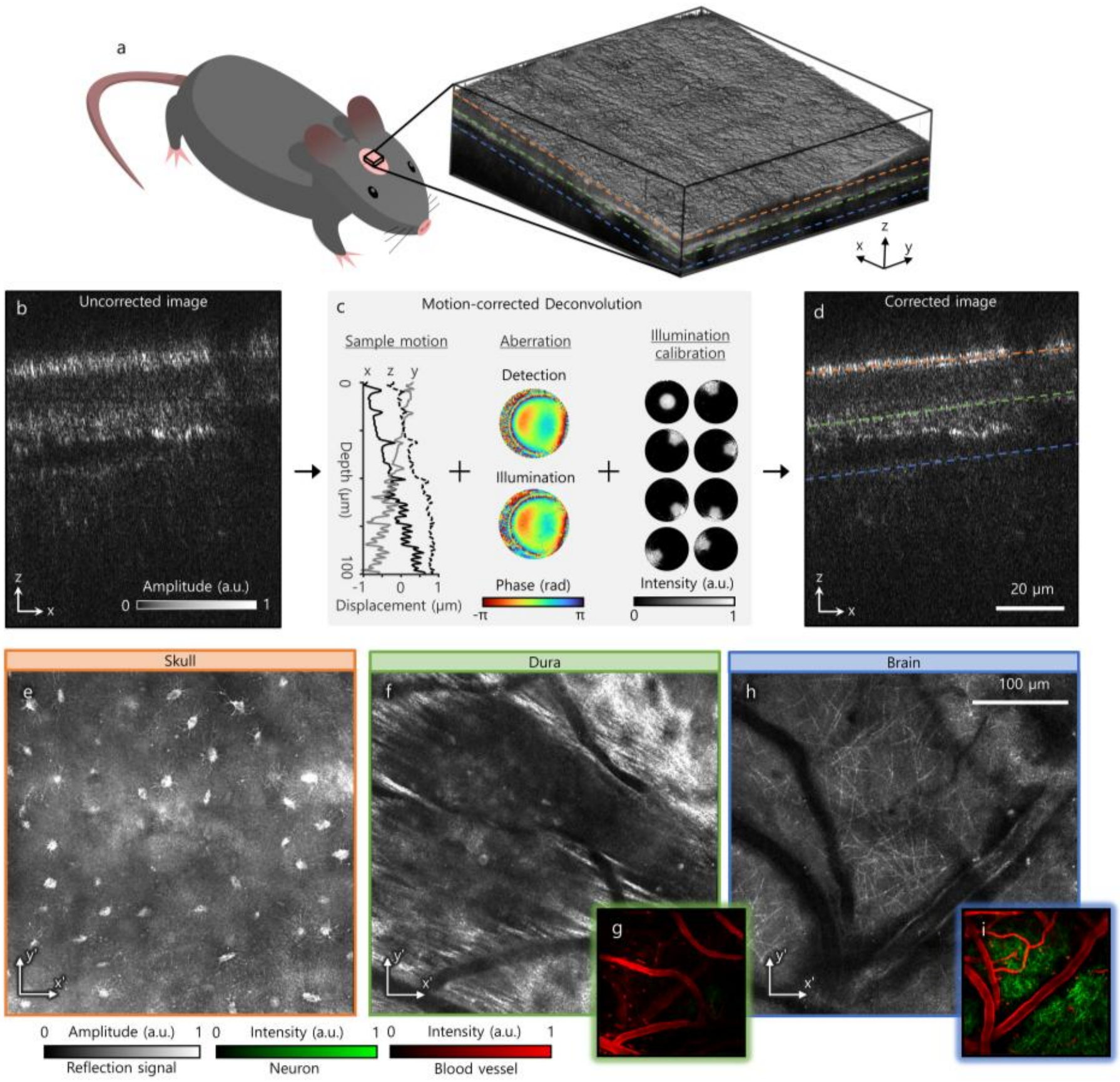

**Figure 6 | *In vivo* volumetric imaging of the mouse brain. a** Reflection tomograms of a mouse brain acquired through a thinned-skull cranial window. **b** Raw reflection data suffers from aberrations and motion artifacts. **c** Joint motion correction, aberration retrieval and illumination self-calibration **d** restore structural contrast and axial resolution. Volumetric reconstructions resolve distinct anatomical layers including **e** the skull, **f** the dura and **h** the cortical tissue. Insets **g** and **i** show corresponding confocal fluorescence images.

## Discussion and conclusion

We developed programmable spatial coherence tomography (PSCT), a fast and robust method for volumetric reflection imaging. PSCT corrects for aberration, stabilizes sample motion and calibrates illumination patterns. The versatility and robustness of PSCT was demonstrated on single cells, large histological tissue, fast sample dynamics and in-vivo imaging.

Compared with plane-wave-based adaptive optical methods[13,14,34], PSCT operates with fewer illumination. In contrast to OCT correction strategies that rely on guide stars or sample-specific assumptions[15,16,35–37], PSCT aberrations are retrieved without modal constraint such as spherical-only correction [24] nor pupil-filtering [38,39]. In addition, the PSCT algorithm also accounts for both illumination and detection aberrations, corrects for illumination alignment and sample motion, providing a more complete calibration than most existing methods.

While PSCT penetration depth is currently limited to approximately 100 µm. We believe that time gating, imaging with longer wavelengths and a lower numerical aperture would improve penetration depth at the cost of a more complex optical setup. Furthermore, the concept and algorithms using programmed spatial coherence can be directly extended to FF-OCT by adding spectrum information in the point spread function.

One of the advantages of PSCT is its simple optical setup not requiring dispersion nor path length matching. This simplicity could enable the development of a unit directly pluggable into the camera port of any microscope. If integrated alongside fluorescence microscopes, the aberration information from PSCT could also help deconvolution of the fluorescence signal. Furthermore, the label-free reflection and dynamic signals are complementary to the high specificity of the fluorescence signal, providing rich complementary information without photobleaching nor toxicity.

By demonstrating that temporal coherence is not required for volumetric reflection imaging, we establish programmable spatial coherence as an alternative design principle to temporal coherence gating and reflection matrix methods, expanding the toolkit available for label-free three-dimensional reflection microscopy.

## Methods

**Acquisition parameters**

For each volumetric acquisition, 8 pupil patterns were programmed on the DMD. For each volume a total of, 1600 camera frames were acquired at 180 fps with a scanning step size of 500 nm, yielding a total acquisition time of 9 seconds per single volume (field of view: 80 × 80 × 100 $\mu m^3$). In various experiments tomograms were stitched to increase the field of view (Colon tissue: 40 × 40 × 1, mouse brain : 5 × 5 × 1, fixed organoids: 3 × 3 × 2, live organoids: 3 × 3 × 3, oocytes: 3 × 3 × 1). The laser power at the sample plane was 30 μW. Volumetric reconstruction was performed in MATLAB R2024b. The field retrieval and motion stabilization time per volume was 4 minutes, while the aberration correction and deconvolution time per volume was 5 minutes on a workstation equipped with an RTX 3090 (Nvidia) graphic processing unit. In the colon tissue image aberration correction was not applied to reduce processing time.

**Colon tissue preparation**

Human colon tissue samples were obtained in accordance with institutional ethical guidelines and approved by the Institutional Review Board of Gangnam Severance Hospital (IRB No. 3-2022-0083). Written informed consent was waived due to the retrospective design and use of anonymized residual specimens without associated clinical data. Two formalin-fixed, paraffin-embedded (FFPE) colon cancer cases were randomly selected by a board-certified pathologist (S.J.S.). Tissue blocks were sectioned to a thickness of 70 μm using a microtome, mounted onto glass slides, deparaffinized in xylene, coverslipped using a commercial mounting medium (Consul-Mount, Epredia), and sealed prior to imaging.

**Mouse oocyte preparation**

All animal procedures were approved by the Institutional Animal Care and Use Committee (IACUC) of KAIST (KA2024-021-v2). Animals were maintained under standard housing conditions with ad libitum access to food and water. Oocytes were obtained from 4-week-old C57BL/6 female mice. Superovulation was induced by intraperitoneal injection of 0.1 mL CARD HyperOva (Cosmo Bio), followed 48 h later by 20 IU human chorionic gonadotropin (hCG; Sigma-Aldrich). Females were euthanized 15–18 h post-hCG, and oviducts were excised to collect cumulus–oocyte complexes (COCs) from the ampullae. COCs were washed in 80 μL drops of human tubal fluid (HTF; Cosmo Bio) under mineral oil and transferred to confocal dishes (SPL Life Sciences). Oocytes were cultured in 20 μL drops of M16 medium (Merck) under mineral oil (2 mL per well), with 7–10 oocytes per drop, and maintained for approximately 24 h prior to imaging.

**Fixed hepatic organoid preparation**

Mouse hepatic organoids were obtained from a commercial source (STEMCELL Technologies) and cultured in HepatiCult™ Organoid Growth Medium (Mouse; STEMCELL Technologies) according to the manufacturer's protocol. Organoids were mechanically dissociated from Matrigel domes, pelleted by centrifugation (300× g, 5 min), and passaged at a 1:3–1:6 split ratio every 5–7 days. For imaging, organoids were fixed in 4% paraformaldehyde, rinsed in phosphate-buffered saline (PBS), and stored in PBS at 4°C until measurement.

**Live hepatocyte organoid preparation**

Primary mouse hepatocytes were isolated and organoids were established as previously described[40]. Hepatocytes were obtained by two-step collagenase perfusion[41], followed by purification using differential centrifugation and Percoll gradient separation[42,43]. Isolated hepatocytes were embedded in Matrigel at a density of 5,000–10,000 cells per well in 24-well plates. After gel polymerization, Hep-Medium[40] was added. The medium consisted of Advanced DMEM/F12 supplemented with HEPES, GlutaMax, penicillin–streptomycin, RSPO1 (100 ng/mL), Wnt surrogate-Fc (0.5 nM), B27 (minus vitamin A), EGF (50 ng/mL), N-acetylcysteine (1.25 mM), gastrin (10 nM), CHIR99021 (3 μM), HGF (25 ng/mL), FGF7 (50 ng/mL), FGF10 (50 ng/mL), A83-01 (1 μM), nicotinamide (10 mM), and Y-27632 (10 μM). Medium was refreshed every three days. Organoids were mechanically fragmented and passaged into fresh Matrigel at a 1:3 ratio after 14 days and subsequently every two weeks.

**Intravital confocal brain imaging**

Thy1-GFP line M mice (Jackson Laboratory, Stock No. 007788) were used for intravital confocal imaging. All procedures were approved by the KAIST Institutional Animal Care and Use Committee (KA2022-057). Mice were anesthetized by intraperitoneal injection of Zoletil (20 mg/kg) and xylazine (11 mg/kg). A thinned-skull cranial window was prepared over the cortical region. Imaging was performed using a commercial intravital confocal microscope (IVM-CMS3, IVIM Technology Inc.) equipped with 488 nm and 640 nm excitation lasers. Emission signals were collected using bandpass filters of 503–558 nm and 672–705 nm, respectively. Images were acquired with a 25× water-immersion objective (CFI75 Apochromat 25XC W, Nikon), yielding a field of view of 483 × 483 μm. To label cerebral vasculature, 25 μg of Alexa Fluor 647-conjugated CD31 antibody (BioLegend) was administered intravenously prior to imaging.

**Acknowledgements**

This work was supported by National Research Foundation of Korea grant funded by the Korea government (MSIT) (RS-2024-00442348, RS-2022-NR068141, RS-2025-00516548, RS-2024-00407383), Korea Institute for

Advancement of Technology (KIAT) through the International Cooperative R&D program (P0028463), the Korean Fund for Regenerative Medicine (KFRM) grant funded by the Korea government (the Ministry of Science and ICT and the Ministry of Health & Welfare) (RS-2024-00332454), and the Samsung Research Funding Center of Samsung Electronics under Grant (SRFC-IT1401-08).

**Conflict of interest**

Y.K.P. is a co-founder and CEO of Tomocube Inc., a company commercializing holographic tomography instruments. P.K. has financial interests in IVIM Technology, a company that commercializes intravital microscopy instruments. Other authors declare no conflicts of interest related to this article.